\magnification=1200 \baselineskip=13pt \hsize=16.5 true cm \vsize=20 true cm
\def\parG{\vskip 10pt} \font\bbold=cmbx10 scaled\magstep2

\centerline {{\it Computer Physics Communications} (2002)}\parG
\centerline {\bbold Simple Dynamics for Broad Histogram Method}\parG
\centerline{Paulo Murilo Castro de Oliveira}\parG
Instituto de F\'\i sica, Universidade Federal Fluminense\par
av. Litor\^anea s/n, Boa Viagem, Niter\'oi RJ, Brazil 24210-340\par
e-mail PMCO @ IF.UFF.BR\par

\vskip 0.5cm\leftskip=1cm\rightskip=1cm 

{\bf Abstract}\parG

        The purpose of this text is: 1) to clarify the foundations of the
broad histogram method, stressing the conceptual differences between it
and reweighting procedures in general; 2) to propose a very simple
microcanonical dynamic rule, yet to be tested by theoretical grounds,
which could provide a good improvement to numerical simulations.\parG


\leftskip=0pt\rightskip=0pt\parG

\vskip 50pt

        The broad histogram method (BHM) [1] is based on the exact [2]
equation

$$g(E) <N_{\rm up}(E)>\, =\, g(E+\Delta{E}) <N_{\rm dn}(E+\Delta{E})>
\,\,\,\, ,\eqno(1)$$

\noindent valid for any system, which is completely general. $g(E)$ counts
the number of states sharing the same energy $E$, and $\Delta{E}$ is a
fixed, arbitrary energy jump. Notation $<Q(E)>$ stands for the
microcanonical average of some quantity $Q$, i.e. its average over all the
above-quoted $g(E)$ states, and only these, uniformly weighted, i.e.

$$<Q(E)> = {1\over g(E)} \sum_{i=1}^{g(E)} Q(S_i)\,\,\,\, .\eqno(2)$$

\noindent The BHM quantities $N_{\rm up}$ and $N_{\rm dn}$ are computed as
follows.

        First, one needs to define a protocol of allowed movements, i.e.
an agreement concerning the changes to be performed on each state $S$
would be considered, tranforming it into another state $S'$. Just to fix
ideas, think about a system of Ising spins, for which one can agree to
perform only single-spin flips, for instance. The only constraint
concerning this protocol is reversibility: if $S \to S'$ is an allowed
movement, then $S' \to S$ is also allowed. For a given state $S$, $N_{\rm
up}(S)$ simply counts the number of such allowed movements increasing its
energy by $\Delta{E}$, the previously defined fixed jump.  Analogously,
$N_{\rm dn}(S)$ counts the number of movements decreasing the energy by
the same amount $\Delta{E}$.

        Is the BHM equation really exact and general for any system?

        This is a very simple matter: the left-hand side of equation (1)
counts the total number of allowed movements from energy level $E$ to
energy level $E+\Delta{E}$; analogously, the right-hand side counts the
total number of reverse movements. Reversibility of the protocol gives a
positive, definitive answer to the above question.

        What has the BHM equation to do with thermodynamics?

        Nothing! It concerns only the energy spectrum of the system under
study, determined exclusively by its Hamiltonian. No hypothesis was made
concerning the particular way the system exchanges energy with the current
environment. Nothing to do with ensembles, equilibrium conditions,
probabilities, transition rates, detailed balances, temperatures, etc. Of
course, $g(E)$ as well as the microcanonical average $<Q(E)>$ of any
quantity is also independent of these thermodynamic concepts. Even the
quoted protocol of movements may have no relation at all with the real
movements which would occur under some particular system-environment
condition. Would the system be simulated on computers, for instance, the
{\bf real} movements performed according to Monte Carlo tossing may be
completely distinct from the {\bf virtual} movements counted by $N_{\rm
up}$ and $N_{\rm dn}$. For instance, one can simulate an Ising system
through cluster algorithms [3], considering only single-spin flips in
order to compute $N_{\rm up}(S)$ and $N_{\rm dn}(S)$ for each visited
state $S$, or vice-versa.

        How to implement the method?

        Suppose one knows the microcanonical averages $<N_{\rm up}(E)>$
and $<N_{\rm dn}(E)>$, as functions of $E$. Then, equation (1) is used to
determine $g(E)$, which is the final goal. Thus, the only remaining task
is to find some way (any way) to determine the $E$-functions $<N_{\rm
up}(E)>$ and $<N_{\rm dn}(E)>$. How to perform this task, of course the
most important issue, is another story to be discussed later on. Once
equation (1) is exact, the numerical accuracy obtained in $g(E)$ depends
exclusively on the corresponding accuracies of the inputs $<N_{\rm
up}(E)>$ and $<N_{\rm dn}(E)>$.

        Note that, until this point, BHM has nothing to do with computer
simulations. Alternative approaches could be adopted in order to obtain
the microcanonical averages $<N_{\rm up}(E)>$ and $<N_{\rm dn}(E)>$
required by BHM. However, I will restrict the discussion hereafter to this
possibility, i.e. to measure these averages from computer simulations.
Even within this restriction, one has many conceivable dynamic rules to
choose.

        What are the advantages over reweighting methods?

        Reweighting methods extract information from the number $V(E)$ of
visits to each energy level, during a previously defined Monte Carlo
dynamics obeying some detailed balance rule. Thus, the relative visitation
frequency to different levels $E$ and $E'$ is previously known, by
construction. One example is the histogram method [4] and its further
developments [5]. Another example is the multicanonical method [6], which
tunes a uniform visitation to all energies. This corresponds to an
acceptation rate from $E$-states to $E'$ proportional to $g(E)/g(E')$:
then, by measuring the acceptation rate actually implemented during the
computer run, one gets $g(E)$ (apart from an irrelevant multiplicative
factor). Other approaches [7] based on this multicanonical idea are also
available. Histogram approaches work similarly, by tuning a
fixed-temperature, canonical distribution of visits. The two main
advantages of BHM over these methods are as follows.

        First, fixed-energy, microcanonical averages do not depend on the
relative visitation to different energy levels. Only a uniform probability
to hit any of the $g(E)$ states sharing the same energy $E$ is required,
no matter what are the visits to other energy levels. Thus, in choosing
the particular dynamic rule to be adopted, one is not restricted to obey
any detailed balance condition, or other similar complications concerning
different energy levels. This freedom could be a big advantage, because
more adequate dynamic rules could be adopted for each particular
application. The only concern is to get a uniform visitation {\bf inside}
each energy level {\bf separately}. Of course, this may not be an easy
task, but at least it is better than to be also forced to obey further
restrictions.

        In particular, any dynamic rule obeying detailed balance
conditions between different energy levels, such as histogram or
multicanonical recipes, also assures a uniform visitation inside each
energy level separately. Thus, any of them could be used within BHM,
simply by measuring $<N_{\rm up}(E)>$ and $<N_{\rm dn}(E)>$ during the
computer run. At the end, instead of the particular prescription of the
original recipe, one can use equation (1) in order to obtain $g(E)$. The
results cannot be worse than those obtained by the original method, once
no further sources of inaccuracies are introduced.

        Second, BHM accuracy is indeed better, when {\bf the same} set of
averaging states obtained by multicanonical recipes, for instance, is
adopted [8]. The reason for that is simple: within reweighting methods,
only the actually tossed movement transforming the current state $S$ into
the next one is considered, whereas BHM takes into account {\bf all
potential} movements one could conceive, starting from $S$. Similarly,
from each state, the infomation extracted by BHM resides on the {\bf
macroscopic} quantities $N_{\rm up}$ and $N_{\rm dn}$, whereas reweighting
methods simply count one more state, $V(E) \to V(E)+1$. Within BHM, $V(E)$
plays no role: it must only be large enough to provide a good statistics.
Moreover, this advantage becomes even larger for larger systems, once
$N_{\rm up}$ and $N_{\rm dn}$ scale with the system size, at least.

        Concluding this part, one possible way to implement BHM is to
adopt, for instance, one of the many available multicanonical dynamics, or
any other obeying some detailed balance condition (perhaps, [9] is the
best one) in order to measure the BHM quantities $<N_{\rm up}(E)>$ and
$<N_{\rm dn}(E)>$. Then, $g(E)$ can be obtained by the exact and general
BHM relation (1), with better accuracy than that corresponding to
exploring the profile of $V(E)$. Obeying detailed balance, the required
uniformity of visits within each energy level is assured. Indeed, this
approach was followed in many recently introduced methods [10], all of
them based on the BHM quantities $<N_{\rm up}(E)>$ and $<N_{\rm dn}(E)>$.
However, this corresponds to an effort larger than one needs, because no
detailed balance between different energy levels is required by BHM.
Another possibility is to profit from that feature by adopting new dynamic
rules which provide a uniform visitation within each energy level
separately. In other words, by relaxing all restrictions concerning the
relative visitation to different energy levels, more efficient computer
simulations could emerge. The corresponding dynamic rules remain to be
invented. Hereafter, I will briefly treat such an attempt, still on the
way.

        How to assure a uniform visitation inside a given energy level?

        My approach to this question is very naive. Consider some ergodic
protocol of allowed movements (again, nothing to do with the BHM protocol
of virtual movements). By tossing random movements within this protocol,
one can construct a Markov chain by simply accepting any new tossed state,
no rejections, no probabilities. Certainly, there is no bias at all: any
state among the whole space of states will be reached according to the
same probability. Thus, for other rules, I conclude that any bias comes
from rejections one is forced to introduce, otherwise the system would
stay forever near the maximum entropy region of the energy spectrum. On
the other hand, for BHM implementations, one does not need complete
uniformity, only within the particular energy level where averages are
measured. My naive idea is to completely avoid rejections {\bf within} the
particular energy level $E$ currently subjected to the averaging process,
although tolerating some rejections out of it. For that, I decided to try
the following simple rule [11]: to accept any tossed movement which keeps
the system inside the window $[E-\delta{E},E+\delta{E}]$, where
$\pm\delta{E}$ is the maximum energy jump allowed by the protocol,
rejecting it otherwise. Although the system is allowed to visit other
energy levels inside this window, averages are taken {\bf only} when the
system visits the central, rejection-free level $E$. In reality, this rule
is not new. It was inspired by a very similar older one [12], introduced
for Ising models within another context.

        Preliminary tests for this dynamics were successful for the
$32\times 32$ square lattice Ising model [11], for which the complete
energy spectrum $g(E)$ is exactly known [13]: I found deviations which
decay by increasing the number $N$ of averaged states inside each energy
level, as $1/\sqrt{N}$, along 6 decades until the maximum $N \approx
10^{10}$ tested. Here, I present further tests for larger $L\times L$
lattices, for which the complete energy spectra are not published in [13],
but the average energy as well as the specific heat is known. Their exact
values at the Onsager temperature [14] are shown in table I, to be
compared with the values obtained by the current approach. These results
seem to indicate that the above-defined dynamics could be approved against
the test of scalability, i.e. it also works well for larger systems. Note
that, even by decreasing $N$ by a factor of 10 for larger lattices, the
deviations seem to remain at the same level. Perhaps this feature is due
to the macroscopic character of $<N_{\rm up}(E)>$ and $<N_{\rm dn}(E)>$.
On the other hand, I am not sure that rejections occurring near the
averaging energy level do not introduce any bias. Visits to the central,
rejection-free level are fed by movements which start at these
neighbouring levels where rejections certainly introduce some biases.
Although not averaged, these biases could reflect themselves into the
central level. Further tests are necessary, with larger lattices, better
statistics and other models. Also, some theoretical work concerning the
sampling uniformity within the central level would be welcome.

        This dynamic rule also presents other advantages. First, one never
compares random numbers with precise probabilities like Boltzmann factors,
which is a delicate matter for other dynamics. Here, random numbers are
used only in order to toss the next movement to be attempted, for instance
the next Ising spin to be flipped or not: the decision to perform it is
deterministic. Then, the quality of the random number generator is not
crucial. Second, the averaging states are not periodically taken, for
instance after each complete lattice sweep, avoiding some extra biases
which could be introduced by this periodicity. On the contrary, a new
averaging state is taken every time the central level is reached, after
some previous random movements. For the square lattice Ising model near
the critical region, this event occurs every 40 movements on average, but
within large fluctuations. Moreover, this number 40 does not depend on the
lattice size, relatively saving computer time for larger lattices.

        Concluding this last part, the dynamic rule presented above
is a promising candidate to be an unbiased microcanonical simulator.
Combined with the broad histogram method, equation (1), it can
contribute to improving numerical simulation studies.

\vfill\eject
{\bf References}\parG

\item{[1]} P.M.C. de Oliveira, T.J.P. Penna and H.J. Herrmann, {\it
Braz.  J. Phys.} {\bf 26}, 677 (1996) (also in Cond-Mat 9610041); for
a review, see P.M.C. de Oliveira, {\it Braz. J. Phys.} {\bf 30}, 195
(2000), (also Cond-Mat 0003300).\par

\item{[2]} P.M.C. de Oliveira, {\it Eur. Phys. J.} {\bf B6}, 111 (1998)
(also in Cond-Mat 9807354); a particular proof, valid only for
single-spin-flip Ising models, is presented in B. Berg and U.H.E.
Hansmann, {\it Eur. Phys. J.} {\bf B6}, 395 (1998) (also in Cond-Mat
9805165, version 2).\par

\item{[3]} R.H. Swendsen and J.-S. Wang, {\it Phys. Rev. Lett.} {\bf
58}, 86 (1987); U. Wolf, {\it Phys. Rev. Lett.} {\bf 62}, 361
(1989).\par

\item{[4]} Z.W. Salzburg, J.D. Jacobson, W. Fickett and W.W. Wood,
{\it J. Chem. Phys.} {\bf 30}, 65 (1959).\par

\item{[5]} G.M. Torrie and J.P. Valleau, {\it Chem. Phys. Lett.} {\bf 28},
578 (1974); B. Bhanot, S. Black, P. Carter and S.  Salvador, {\it Phys.
Lett.} {\bf B183}, 381 (1987); M. Karliner, S.  Sharpe and Y. Chang, {\it
Nucl. Phys.} {\bf B302}, 204 (1988); R.H. Swendsen, {\it Physica} {\bf
A194}, 53 (1993), and references therein.\par

\item{[6]} B.A. Berg and T. Neuhaus, {\it Phys. Lett.} {\bf B267}, 249
(1991); B.A. Berg, {\it Int. J. Mod. Phys.} {\bf C4}, 249 (1993).\par

\item{[7]} A.P. Lyubartsev, A.A. Martsinovski, S.V. Shevkunov and P.N.
Vorontsov-Velyaminov, {\it J. Chem. Phys.} {\bf 96}, 1776 (1992); E.
Marinari and G. Parisi, {\it Europhys. Lett.} {\bf 19}, 451 (1992); J.
Lee, {\it Phys. Rev. Lett.} {\bf 71}, 211 (1993); B. Hesselbo and R.B.
Stinchcombe, {\it Phys. Rev. Lett.} {\bf 74}, 2151 (1995).\par

\item{[8]} A.R. de Lima, P.M.C. de Oliveira and T.J.P. Penna, {\it J.
Stat. Phys.} {\bf 99}, 691 (2000) (also in Cond-Mat 0002176).\par

\item{[9]} F. Wang and D.P. Landau, Cond-Mat 0107006.\par

\item{[10]} J.-S. Wang, T.K. Tay and R.H. Swendsen, {\it Phys. Rev.
Lett.} {\bf 82}, 476 (1999); J.-S. Wang and L.W. Lee, Cond-Mat
9903224; M. Kastner, J.D. Munoz and M. Promberger, {\it Phys. Rev.}
{\bf E62}, 7422 (2000) (also in Cond-Mat 9906097); R.H. Swendsen,
J.-S. Wang, S.-T. Li, C. Genovese, B.  Diggs and J.B. Kadane,
Cond-Mat 9908461; J.-S. Wang, {\it Comput. Phys.  Comm.} {\bf
121-122}, 22 (1999); Cond-Mat 9909177.\par

\item{[11]} P.M.C. de Oliveira, {\it Braz. J. Phys.} {\bf 30}, 766
(2000) (also in Cond-Mat 0101171).\par

\item{[12]} K.-C. Lee, {\it J. Phys.} {\bf A23}, 2087 (1990).\par

\item{[13]} P.D. Beale, {\it Phys. Rev. Lett.} {\bf 76}, 78 (1996).\par

\item{[14]} see, for instance, J. Salas, Cond-Mat 0009054.\par

\vfill\eject
{\bf Table}\parG

\parG\settabs 6\columns
\+\hfill $L$\hfill&
\hfill 32\hfill&
\hfill 46\hfill&
\hfill 62\hfill&
\hfill 90\hfill&
\hfill 126\hfill&\cr
\parG

\+\hfill $e$ (BHM)\hfill&
\hfill 0.141586\hfill&
\hfill 0.143063\hfill&
\hfill 0.143937\hfill&
\hfill 0.144723\hfill&
\hfill 0.145205\hfill&\cr

\+\hfill exact\hfill&
\hfill 0.141585\hfill&
\hfill 0.143064\hfill&
\hfill 0.143937\hfill&
\hfill 0.144718\hfill&
\hfill 0.145212\hfill&\cr
\parG

\+\hfill $C$ (BHM)\hfill&
\hfill 1.846663\hfill&
\hfill 2.028195\hfill&
\hfill 2.175960\hfill&
\hfill 2.362453\hfill&
\hfill 2.528842\hfill&\cr

\+\hfill exact\hfill&
\hfill 1.846768\hfill&
\hfill 2.027854\hfill&
\hfill 2.176425\hfill&
\hfill 2.361582\hfill&
\hfill 2.528522\hfill&\cr

\parG\parG
\item{Table I} 
Average energy per site ($e$) and specific heat ($C$) for $L\times L$
square lattice Ising model at the Onsager temperature. The number of
averaging states per energy level is $N = 10^{10}$ for $L = 32$, 46 and
62, or $N = 10^9$ for $L = 90$ and 126.\par

\bye